\documentstyle[12pt]{article}
\begin{document}
\setlength{\textheight}{8.54truein}
\setlength{\textwidth}{6.0truein}
\voffset=-1in
\leftmargin=0.5in
\rightmargin=0.5in
\def\REF{\hangindent=4em \hangafter=1 \noindent}
\def\refrule{\hbox to 3pc{\leaders\hrule depth-2pt height 2.4pt\hfill}}
\def\cmax{ {C_{\rm max}} }
\def\cmin{ {C_{\rm min}} }
\def\ccut{ {C_{\rm cut}} }
\def\ccutp{{C_{\rm cut}^{'}}}
\def\ccma{( C_{\rm max} /C_{\rm cut})_{1024}}
\def\ccmb{( C_{\rm max} /C_{\rm cut})_{64}}
\def\deltat{ \Delta t_{90} }
\def\vmax{  V/V_{\rm max} }
\def\vvmax{ \langle V/V_{\rm max} \rangle }
\def\cost{ \langle \cos\theta \rangle}
\def\sinb{ \langle \sin^2 b \rangle}
\def\ccx{\cmax/\cmin}
\def\etal{{\it etal.}}
\def\sles{\lower2pt\hbox{$\buildrel {\scriptstyle <} \over
{\scriptstyle\sim}$}}
\parskip=10pt
\centerline{\large \bf On the Bimodal Distribution of Gamma-Ray Bursts}
\vspace{0.5cm}
\centerline{Shude Mao, Ramesh Narayan and Tsvi Piran\footnote{
Permanent address: Racah Institute for Physics, The Hebrew University,
Jerusalem, Israel}}
\vspace{0.5cm}
\centerline{Harvard-Smithsonian Center for Astrophysics, 60 Garden Street,
Cambridge, MA 02138}

\vspace{0.5cm}
\centerline{\it Received .........;}
\vspace{0.5cm}

\centerline{ABSTRACT}
\vspace{0.3cm}

Kouveliotou et al. (1993) recently confirmed that gamma-ray bursts are
bimodal in duration. In this paper we compute the statistical properties of
the short ($\le 2$~s) and long ($>2$~s) bursts using a method of analysis
that makes no assumption regarding the location of the bursts, whether in the
Galaxy or at a cosmological distance. We find the 64 ms channel on BATSE
to be more sensitive to short bursts and the 1024 ms channel is more sensitive
to long bursts. We show that all the currently available data are consistent
with the simple hypothesis that both short and long bursts have the same
spatial
distribution and that within each population the sources are standard
candles. The rate of short bursts is $\sim 0.4$ of the rate of long bursts.
Although the durations of short and long gamma-ray bursts span several orders
of magnitude and the total energy of a typical short burst is smaller
than that of a typical long burst by a factor of $\sim 20$,
surprisingly the peak luminosities of the two kinds of bursts are equal
to within a factor of $\sim 2$.

\newpage

\section{Introduction}

The Burst and Transient Source Experiment (BATSE) on board the
Compton Gamma-Ray Observatory has demonstrated that the distribution
of gamma-ray bursts is isotropic over the sky and bound in the radial
direction (Meegan et al. 1992).  This strongly indicates that the
sources of the bursts are located either at cosmological distances
(Paczy\'nski 1991, Dermer 1992, Mao \& Paczy\'nski 1992, Piran 1992)
or in an extended Galactic halo (Li \& Dermer 1992, Duncan, Li, \&
Thompson 1992).  Many models have been proposed for the cosmological
scenario (e.g., Eichler, et al., 1989, Piran, Narayan, \& Shemi, 1992,
Narayan, Paczy\'nski, \& Piran 1992, Rees \&
M\'esz\'aros 1992, Usov 1992, Woosley 1993) as well as for the
extended halo scenario (Fabian \& Podsiadlowski 1993).  For a
recent review on gamma-ray bursts, see  Paczy\'nski (1992).

Recently Kouveliotou et al. (1993) showed that the distribution of
durations of gamma-ray bursts is bimodal.  In terms of the parameter
$\deltat$, which is the time interval during which the integrated
counts of a burst go from 5\% to 95\% of the total integrated counts,
the bursts seem to separate cleanly into two distinct groups, with
the transition occurring around $\deltat \approx 2$ s.  This confirms
similar indications from earlier experiments (Cline \& Desai 1974,
Norris et al. 1984, Dezalay et al. 1992, Hurley 1992), and is the first
compelling evidence for distinct sub-classes of gamma-ray bursts.  (We refer
here to classical gamma-ray bursts, which represent the bulk of the
bursts, and do not discuss the class of soft repeaters.)

An obvious and natural question that arises from this result is: what
is the relation between the two kinds of bursts?  One possibility is
that they represent two distinct types of sources.  For instance, it
is even conceivable that one class of sources is cosmological and
that the other is in the Galactic halo.  Alternatively, both types of
bursts may come from a common source, and the differences may arise
merely from variations in the initial conditions of the sources prior
to the bursts, or changes in the environment or the viewing angle. We
attempt to shed
some light on this question by carrying out a statistical comparison
of the properties of the two kinds of bursts.  We describe the detector
channels on BATSE and some selection effects in \S 2, present the main data
analysis in \S 3, and discuss the implications of the results in \S
4.

\section{Detector Channels and Selection Effects}

The burst catalog available in the public domain contains a list of
all gamma-ray bursts that triggered the BATSE detectors between April
1991 and March 1992 (Fishman et al. 1993).  The catalog provides the
angular positions for 260 bursts, the ratio $\cmax/\cmin$ for 241
bursts, where $\cmax$ is the maximum count rate and $\cmin$ is the
detection threshold, the fluences and peak count rates for 260 bursts,
and durations for 220 bursts.

The trigger mechanism and various selection effects of BATSE have been
explained in detail in Fishman (1992).  We will repeat the essentials
here.  The BATSE on-board software tests for bursts by comparing count
rates on eight large-area detectors to the threshold levels
corresponding to three separate time intervals: 64 ms, 256 ms, and
1024 ms.  A burst trigger occurs if the count rate is above the
threshold in two or more detectors simultaneously.  The thresholds are
set by command to a specified number of standard deviations above the
background (nominally 5.5 $\sigma$), and the average background rate
is computed every 17 s.
Due to an unknown technical reason, many bursts have ``undetermined''
$\cmax/\cmin$ in the 256 ms channel.  We therefore ignore this channel
in what follows.

For a variety of reasons, BATSE has a variable background as a
function of time, so that the detection threshold $\cmin$ does not
remain constant.  Moreover, there are periods corresponding to
``overwrites'' when the sensitivity of the detectors is greatly
reduced.  These effects make an analysis of the $\cmax /\cmin$ data
somewhat difficult.  In order to have a more uniform sample to work
with, we select a constant threshold $\ccut$, and prune the data so as
to include only those bursts which satisfy both of the following
criteria:

\begin{itemize}
\begin{enumerate}
\item $\cmin \le \ccut$,
\item $\cmax \ge \ccut$.
\end{enumerate}
\end{itemize}

\noindent
The resulting database corresponds to those bursts which would have
been found by a detector that (i) had a constant threshold of $\ccut$,
and (ii) was turned on at those times when the real detectors on
BATSE had $\cmin \le \ccut$, and was turned off whenever the BATSE
sensitivity was poorer than $\ccut$.  By cutting the data in this
fashion, we are guaranteed to have a sample of bursts with a constant
detection threshold and a uniform selection bias.  Of course, in the
process we lose a few bursts, which causes loss of statistical
accuracy.  To minimize this, we select $\ccut$ such that the number of
bursts retained in the database is maximized.  This leads to the
choice $\ccut=71$ counts for the 64 ms channel and $\ccut=286$ counts
for the 1024 ms channel.  Fortunately, only a few bursts are
eliminated for these choices of $\ccut$; moreover, the excluded bursts
are mostly those that are labeled ``overwrites'', and would have been
eliminated in any case.

Following Kouveliotou et al. (1993) we define ``short bursts'' as
having $\deltat \le 2$~s, and ``long bursts'' as having $\deltat >
2$~s.  Before going into the main analysis, which we discuss in the
next section, we explain first some selection effects associated with
the different sensitivities of the 64 ms and 1024 ms channels to the
two kinds of bursts.  Fig. 1 shows $\ccma$ in the 1024 ms channel vs.
$\ccmb$ in the 64 ms channel for the short and long bursts.  It is
apparent that the 1024 ms channel is more sensitive to long bursts,
while the 64 ms channel is more sensitive to short bursts.  Both of
these effects are quite natural, as we now show.

The noise in the background counts increases, generally, as the square
root of the integration time.  Therefore the noise is expected to be 4
times greater in the 1024 ms channel than in the 64ms channel.  Now,
if a burst has a broad luminosity maximum extending over a time
interval greater than 1024 ms then there will be 16 times more signal
counts in the 1024 ms channel than in the 64 ms channel, and the
signal-to-noise ratio will be 4 times greater.  Long bursts are likely
to display this behavior.  On the other hand, if a burst is extremely
narrow, with a duration less than 64 ms, then the number of signal
counts will be the same in both channels.  The most extreme short
bursts will correspond to this limit.  Based on this argument, we see
that the ratio $\ccma$ to $\ccmb$ for the two channels must satisfy $$
{1 \over 4} <{ \ccma
\over \ccmb } < 4,  \eqno (1)
$$
with short and long bursts tending towards the lower and upper limit
respectively.

Comparing the ratio in eq (1) for the long burst  bursts, we find
$$
R_1=\left\langle { \ccma \over \ccmb } \right\rangle_{\rm long} = 2.5
\pm 0.64 , \eqno (2)
$$ where the error estimate reflects the width of the distribution.
Since $R_1$ is not very different from the maximum value of 4, we
conclude that, to a first approximation, the long bursts tend to have
broad nearly constant luminosity profiles.  For a more detailed
description of the intensity statistics,
we note that the maximum luminosity in the 64 ms channel is greater by
a factor $\sim 4/2.5=1.6$ than the luminosity in the 1024 channel.
This may be interpreted as evidence for a ``fractal'' behavior in the
burst luminosity, $$
\bar L (\Delta t)
\propto \Delta t^{-0.2},\eqno (3)
$$
where $\bar L(\Delta t)$ represents the maximum luminosity of a
burst as measured with a time constant of $\Delta t$.

In the case of the short bursts we find
$$
R_2=\left\langle {
\ccma \over \ccmb } \right\rangle_{\rm short} = 0.76 \pm 0.48, \eqno
(4)
$$
which shows that the 64 ms channel is more sensitive than the
1024 ms channel.  This is almost entirely because the longer channel
dilutes the signal.  As clear evidence of this effect we note that
there is a strong correlation between the widths $\deltat$ of the
short bursts and the quantity ${\ccma / \ccmb }$ (the correlation
coefficient is 0.49).

\section{Data Analysis}

In this section we carry out a comparison among various samples of
bursts.  We work primarily with three samples:
\begin{itemize}
\item sample $s_{64}$ consisting
of 40 short bursts which were detected in the 64 ms channel,
\item
sample $l_{1024}$ consisting of 113 long bursts detected in the 1024
ms channel, and
\item sample $l_{1024,64}$ consisting of 71 long bursts
detected in both the 1024 ms and 64 ms channels.
\end{itemize}

In comparing different samples, the key observational data we use are
the distributions of $\vmax$ of the samples, where $\vmax = (\cmax
/\ccut)^{-3/2}$ (Schmidt et al. 1988). Our analysis is based on a simple
hypothesis,
namely that all three populations of bursts have the same underlying
spatial distribution, and that within each population the sources are standard
candles. On this hypothesis, any apparent differences between the
samples are just because they are viewed to different distances (or
depths) on account of differences in the source luminosity and/or
detector sensitivity.  Our main result is that all the available data
are consistent with this idea.

In our analysis, we make use of the fact that the brighter bursts
agree with a homogeneous Euclidean distribution, which is
characterized by cumulative counts varying linearly as $\vmax$ and a
mean $\vvmax$ equal to 0.5.  In contrast, the fainter bursts deviate
significantly from such a distribution.  Based on this observation, we
make the following reasonable assumptions:

\noindent
(1) We assume that $\vvmax$ decreases monotonically with increasing
depth of a sample.  Therefore, if we compare two samples of bursts
(with the same spatial distribution by hypothesis), we can say that
the sample with the smaller value of $\vvmax$ corresponds to a greater depth or
distance.  Conversely, if two samples have the same value of $\vvmax$
we say that they correspond to the same distance.  As a further check,
when $\vvmax$ of two samples agree, we compare their full $\vmax$
distributions by means of the Kolmogorov-Smirnov test (K-S test), and
thus investigate whether or not the two populations do indeed have
a common spatial distribution.

\noindent
(2) If two samples have different $\vmax$ distributions and different
mean $\vvmax$, but we suspect that they have intrinsically the same
spatial distribution, then we can bring the two samples into agreement
by increasing $\ccut$ for the deeper population.  In effect, we
artificially reduce the sensitivity of the detector corresponding to
the deeper sample so that its sensitivity becomes equal to that of the
shallower sample.  We can think of this operation equivalently as
reducing the luminosity of all bursts in the deep sample by a constant
factor.  Of course the procedure is meaningful only if it leads to
agreement in the mean $\vvmax$ and also in the shapes of the $\vmax$
distributions, as discussed in (1) above.

\noindent
A point that we would emphasize is that our method of analysis is virtually
model-free and applies regardless of whether bursts are Galactic or
cosmological.

To illustrate the method we consider the two samples of long bursts,
$l_{1024}$ and $l_{1024,64}$.  The discussion in \S 2 showed that the
1024 ms channel is more sensitive to long bursts than the 64 ms
channel by a factor of 2.5 (eq 2).  We therefore expect $l_{1024}$ to
correspond to a deeper sample than $l_{1024,64}$.  This is confirmed
by the mean $\vvmax$ values, which are $0.29 \pm 0.027$ for $l_{1024}$
and $0.38 \pm 0.034$ for $l_{1024,64}$.  Both groups deviate
significantly from a uniform distribution in Euclidean space, but the
deviation is much larger for $l_{1024}$.  We can now artificially
bring the two populations to the same distance by increasing $\ccut$
for the $l_{1024}$ sample by a factor of 2.5.  On doing this we find
that the value of $\vvmax$ for $l_{1024}$ becomes $0.37 \pm 0.035$,
which is nearly equal to the $\vvmax$ of the $l_{1024,64}$ sample,
exactly as expected.  Furthermore, the two cumulative $V/V_{max}$
distributions agree very well with each other after this distance
correction has been done.  The K-S probability (for a worse fit than
the one obtained) is 86\%, which is excellent.  These calculations
show that the $l_{1024}$ and $l_{1024,64}$ samples do have the same
spatial distribution, with the former being a deeper sample than the
latter by a factor of $\sqrt {2.5}$ in luminosity distance.  This
is no surprise since the two samples have a large number of
bursts in common, and moreover we know that there is a good
correlation between the signals in the 1024 ms and 64 ms channels for
long bursts (cf Fig. 1).  The test however demonstrates the validity
of the method.

We next proceed to the more interesting test of comparing the short and
long bursts.  The average $\langle V/V_{max} \rangle $ of the $s_{64}$
sample is $0.31 \pm 0.042$, as compared to $0.29 \pm 0.027$ for
$l_{1024}$, which indicates that the two samples correspond to nearly
the same distance.  We now vary $\ccut$ of the $l_{1024}$ sample and
$s_{64}$ sample individually so as to find the range of values over
which the $\vvmax$ values of the two samples agree to within $\pm
1\sigma$ (see Fig. 2). We find that
$$ R_3 = \left({\ccutp \over\ccut}\right )_{l_{1024}}
\left({\ccut \over \ccutp}\right )_{s_{64}} =
1.4^{+1.3}_{-0.6}, \eqno (5) $$
where the value $R_3=1.4$ corresponds to the case when the two $\vvmax$
values are exactly equal..
As defined here, $\sqrt{R_3}$
represents the ratio of the limiting distances of the $l_{1024}$ and $s_{64}$
samples.  We should mention in passing that there is a second region
of good fit around $R_3\sim 6$, but the number of bursts that survive
the cut for this comparison is so small that we do not find the
solution convincing. Even the primary solution given in eq (5) suffers
to some extent from the limited number of bursts in the two samples.

We now test whether or not the $s_{64}$ and $l_{1024}$ populations are
really consistent with the same spatial distribution.  For this we do
a K-S test to compare the two distributions of $\vmax$ (see Fig. 2).
We see that the test indicates fairly convincingly that the two
populations do have the same spatial distribution.  For instance, the
K-S probability is 42\% if we keep the two $\ccut$ values unchanged
($R_3=1$) and 50\% if we multiply $\ccut$ for the 1024 ms channel by
the optimum factor of 1.4 to obtain equality of $\vvmax$.  To
illustrate the quality of the agreement, we show in Fig. 3 the $\vmax$
distributions corresponding to the case when $R_3=1$.  Note how much
better the two distributions agree with each other than with the
diagonal line which represents the homogeneous Euclidean model.  The
probability that either of the observed samples is drawn from the
Euclidean distribution, is vanishingly small, $< 10^{-4}$ according
to the K-S test.

Finally, we compare the $s_{64}$ and $l_{1024,64}$ samples.
In analogy with eq (5) we define a corresponding ratio $R_4$
for this comparison,
$$
R_4\equiv \left({\ccutp \over\ccut}\right )_{s_{64}}
\left({\ccut \over \ccutp}\right )_{l_{1024,64}}. \eqno (6) $$
For consistency with the two previous comparisons, we expect $$
R_4={R_1\over R_3}= 1.8^{+2}_{-1}.\eqno (7) $$ To check this, we
change $\ccut$ for the $s_{64}$ sample from 71 counts to
$\ccutp=1.8\times 71$ counts and compare the modified $s_{64}$ sample
with $l_{1024,64}$.  The mean $\vvmax$ values of the two samples are
$0.41\pm 0.047$ for $s_{64}$ and $0.38\pm 0.034$ for $l_{1024,64}$,
showing excellent agreement.  Further, the K-S test gives a high
probability of 46\%, again in good agreement.  We thus conclude that
the $s_{64}$ and $l_{1024,64}$ samples are consistent with a common
spatial distribution, and that the former is deeper than the latter by
about $\sqrt{1.8}$ in luminosity distance.  (As an aside we mention
that, corresponding to the second solution $R_3\sim 6$ mentioned
earlier, there is an indication of a solution at $R_4\sim 0.5$, but
the reduction in counts in these comparisons is somewhat severe and we are
inclined not to take these solutions seriously.)

A very interesting feature of the comparison discussed in the previous
paragraph is that $R_4$ directly represents the ratio of the
luminosities of the long and the short bursts {\it in the same
detector}, viz. the 64 ms channel.  Since the 64 ms channel
corresponds to the shortest time interval used in the triggers,
it provides the closest approximation to the instantaneous luminosity
of a burst.  We thus
conclude that {\it the maximum instantaneous luminosities of the short
and long bursts are nearly equal}, to within a factor $\sim 2$.  In
fact, the difference in the luminosities may be even less than the
value indicated in (7).  This is because, in the cosmological
scenario, the K-correction (cf.  Piran 1992; Mao \& Paczy\'nski 1992)
depends on the spectral index of the gamma-ray bursts.  Although the
spectral indices of bursts have large variations, there is some
indication that the short bursts are systematically harder than the
long ones (Dezalay et al. 1992; Kouveliotou et al. 1993). The effects
of this will be to bring the luminosity ratio even closer to unity.
Utilizing the fact that we have $\cmin=71$ counts in the 64ms channel,
we can {\it roughly} estimate the peak luminosity of bursts
by assuming a power law spectrum (see eq 10 below):
$$L_{\rm
 peak}\sim 2 \times 10^{43} (D_{\rm max}/{\rm Mpc})^2 ~ {\rm erg ~ s^{-1}},
\eqno(8)$$
where $D_{\rm max}$ is the maximum distance to which the bursts can be
detected in the 64 ms channel.

An important consequence of the comparisons carried out above
is that we obtain the relative depths of the
short and long burst samples. We can therefore estimate the
relative number densities in space of the two kinds of bursts.
We find
$$ {n_S\over n_L} \sim 0.4^{+0.4}_{-0.2}. \eqno (9) $$
We should however mention one caveat, namely
that BATSE may have missed some very short bursts with durations
smaller than 64 ms because of dilution.  If this is a substantial
effect, then the above ratio may be higher, possibly closer to unity.

We now estimate the ratio of the total energy outputs in the short and
long bursts. We first define an effective duration $\Delta t_{\rm eff}$ as
the duration of a burst would have if it had a constant count rate $\cmax$
and the same fluence, i.e.,
$$
\Delta
t_{\rm eff} = {S \over \langle E \rangle ~ \cmax}, ~~~  \langle E \rangle
\equiv
{\int_{E_1}^{E_2} E^{-\alpha+1} dE \over \int_{E_1}^{E_2} E^{-\alpha}
dE}, \eqno (10)
$$
where $S$ is the fluence in units of $\rm erg ~ cm^{-2}$ in the
energy range 50--300 keV (which coincides with the trigger energy
range), $\langle E \rangle$ is the mean energy in the energy range of
50-300 keV, $\cmax$ is the maximum count rate in units of
$\rm photons ~ cm^{-2} ~ s^{-1}$, and we assume a power law photon number
distribution $n(E)~dE \propto E^{-\alpha}~dE$.  Adopting $\alpha=2$
(Schaefer et al. 1992), we find that $ \langle \Delta t_{\rm eff}
\rangle \approx 0.4~{\rm s},~12.5~{\rm s}$
for the short and long bursts respectively.  Combining this with the
luminosity ratio $R_4$, we find the total energy ratio is $E_L/E_S
\approx 20$.  It is quite remarkable that the short and long bursts
differ by such large factors in their durations and total energy
outputs, but yet are so similar in their maximum luminosities.
Finally, for completeness, we mention that the short and long
bursts are both individually consistent with perfect isotropy.

\section{Discussion}

This paper has been motivated by the recent discovery of Kouveliotou et al.
(1993) that gamma-ray bursts consist of two distinct subclasses,
namely short and long bursts.  The main aim of our investigation is to
test the simple hypothesis that both populations of bursts have the
same underlying space distribution and that within each population
the sources are standard candles.  Our conclusion is that all the
available data are consistent with this hypothesis.  Even though
the distributions of $\cmax /\ccut$ for the short and long bursts
sometimes appear to be different in certain detector channels, we are
always able to bring the two populations into agreement by modifying
the detector sensitivity in one or the other sample so as to reduce
the two samples to the same depth or distance.  When we do this, not
only do the mean $\vvmax$ values agree, but also the full distributions of
$\vmax$ agree when compared by means of the K-S test.  Of course,
these calculations do not prove our basic hypothesis, particularly
since we are hampered by the small number of bursts in the samples
but they make the idea quite plausible.  We therefore feel that it is unlikely
that one type of bursts (say long) is cosmological while the other
arises in the halo, or that one is in the halo and the other in the
disk (Smith \& Lamb 1993). It would be too much of an accident for the two
populations to have the same $V/V_{\rm max}$ distributions.
Instead, we favor models where the two kinds of
bursts arise in the same source.  In this case, the large difference in
durations between the short and long bursts may be caused by variations
in the initial conditions or in the environment of the source or due to
differences in the viewing angle.

Once we accept that the two kinds of bursts arise from a common source
population, we are able to estimate the relative luminosities and
number densities of the two populations.  Surprisingly, we find that
both the short and long bursts have the same peak luminosity, $L_{\rm
peak}\sim 2 \times 10^{43} (D_{\rm max}/{\rm Mpc})^2 {\rm erg ~ s^{-1}}$,
to within a factor of two.  The equality of the two peak luminosities is quite
remarkable when we consider that the durations of the short and long
bursts differ by $\sim 50$ and their total energy outputs differ by
$\sim 20$.  Incidentally, our estimate of $L_{\rm peak}$ corresponds
to the 64 ms channel.  On smaller timescales, the peak luminosity will
be larger, though not by a significant factor (see eq 3). Further, we estimate
the number density of the short bursts to be $\sim 0.4$ times that of
the long bursts.  If the difference between short and long bursts is due
to viewing angle, this ratio gives an estimate of the relative solid
angles associated with the two kinds of bursts.

An important point that we would stress is that these results are
obtained in an essentially model-independent way.  For instance, we do
not need to make any assumption on whether the sources are at
cosmological distances or in the Galaxy.

The constancy of luminosity in two classes of bursts which differ so
much in their other properties might provide a clue to the physical
origin of the bursts.  Unfortunately, it is virtually impossible to
explain the result using the most widely used limiting luminosity in
astrophysics, namely the Eddington limit.  If we assume that the
sources are not dynamically expanding (but see Piran and Shemi, 1993)
and take a fixed opacity (e.g. electron scattering), the Eddington
limit is proportional to the mass of the source, which in turn is
limited by the variability timescale, $\delta t$, i.e., $L_{\rm Edd} < 1.3
\times 10^{40} (\delta t/{\rm ms}) ~ {\rm erg ~ s^{-1}}$.  Taking $\delta t
\sim 10$ ms, the characteristic burst luminosity $L_{\rm peak}$
that we have obtained is marginally consistent with the Eddington
luminosity limit for sources of mass $\sim 10^3 M_\odot$ located at
distances $\sim 100 \rm kpc$ in the Galactic halo.  Smaller masses are
ruled out by the observed isotropy of the bursts, while larger masses
are ruled out by the variability argument.  In fact, Bhat et al. (1993) claim
to see variability down to $200~\mu$s, which rules out even the $10^3
M_\odot$ scenario.

Another robust limiting luminosity is $L_{\rm max} = c^5/G = 4\times
10^{59} ~{\rm erg ~ s}^{-1}$, which is an absolute upper limit for any
source, corresponding to the emission of the entire rest mass within
a gravitational light crossing time.  If we identify $L_{\rm peak}$ with
this limit, then we obtain the luminosity distance to the sources
to be $\sim 10^8 ~ {\rm Mpc}$, corresponding to a redshift of
$\sim 10^4$, in an $\Omega =1$ Friedmann universe. This is far
too large for any known model.

The fact that it is not easy to come up with a physical explanation
for the existence of a characteristic peak luminosity for gamma-ray
bursts implies that, if the effect is real, it may provide an
important and vital clue for understanding the origin of the bursts.
Unfortunately, as we have tried to stress, the results have only modest
statistical significance at this point.  It would be very interesting to
repeat the analysis with the complete database of bursts detected by
BATSE.

We acknowledge helpful discussions with Bohdan Paczy\'nski.  This
work was supported in part by NASA grant NAG 5-1904.

\newpage

\section{References}
\parskip=2pt

\REF{Bhat, P. N., Fishman, G. J., Meegan, C. A., Wilson, R. B.,
Brock, M. N., \& Paciesas, W. S. 1992, Nature, 359, 217}

\REF{Briggs, M., 1993, ApJ, in press}

\REF{Cline, T. L., \& Desai, U. D. 1974, in the Proc. 9th ESSLAB Symposium,
37--45 (ESRO, Noodrwijk)}

\REF{Dermer, C. D., 1992, Phys. Rev. Lett., 68, 1799}

\REF{Dezalay, J-P., Barat, C., Talon, R., Sunyaev, R., Terekhov, O., \&
Kuznetsov, A. 1992, in AIP Conference
Proceedings 265: Gamma-Ray Bursts, ed. W. S. Paciesas \& G. J. Fishman, p. 304}

\REF{Duncan, R. C., Li, H., \& Thompson, C. 1992, in the Conference
Proceedings of the Compton Symposium (St. Louis, Oct. 15--17), in press}

\REF{Eichler, D., Livio, M., Piran, T., \& Schramm, D. N. 1992, Nature,
340, 126}

\REF{Fabian, A. C., \& Podsiadlowski, P. 1993, MNRAS, in press}

\REF{Fishman, G. J. 1992, in the Compton Observatory Science
Workshop, ed. C. R. Shrader, N. Gehrels, \& B. Dennis (NASA Conference
Publication 3137)}

\REF{Fishman, G. J. et al. 1993, ApJS, in preparation}

\REF{Harding, A. K. 1991, Physics Reports, 206, 327}

\REF{Higdon, J. C., \& Lingenfelter, R. E. 1990, ARA\&A, 28, 401}

\REF{Hurley, K. 1992, in AIP Conference Proceedings 265: Gamma-Ray
Bursts, ed. W. S. Paciesas \& G. J. Fishman, p. 3}

\REF{Kouveliotou, C. et al., ApJL, in preparation}

\REF{Li, H., \& Dermer, C. D. 1992, Nature, 359, 514}

\REF{Mao, S., \& Paczy\'nski, B. 1992, ApJ, 388, L45}

\REF{Meegan, C. A. et al. 1992, Nature, 355, 143}

\REF{Narayan, R., Paczy\'nski, B., \& Piran, T. 1992, ApJ, 395, L83}

\REF{Norris, J. P., Cline, T. L., Desai, U. D., \& Teegarden, B. J. 1984,
Nature, 308, 434}

\REF{Paczy\'nski, B. 1991, Acta Astronomica, 41, 157}

\REF{\refrule. 1992, presented at the TEXAS/PASCOS conference
at Berkeley, Dec. 1992}

\REF{Piran, T. 1992, ApJ, 389, L45}

\REF{Piran, T., Narayan, R., \& Shemi, A. 1992, in AIP Conference
Proceedings 265: Gamma-Ray Bursts, ed. W. S. Paciesas \& G. J. Fishman, p. 149}

\REF{Piran, T., \& Shemi, A. 1993, ApJ, 403, L67}

\REF{Rees, M., \& M\'esz\'aros, P. 1992, MNRAS, 258, 41P}

\REF{Schaefer, B. E. et al. 1992, ApJ, 393, L51-L54}

\REF{Schmidt, M., Higdon, J. C., \& Hueter, G. 1988, ApJ, 329, L85}

\REF{Smith, I. I., \& Lamb, D. Q. 1993, ApJ, in press}

\REF{Usov, V. V. 1992, Nature, 357, 472-475}

\REF{Woosley, S. E. 1993, ApJ, in press}

\newpage

\section{Figure Captions}

\noindent
{\bf Fig. 1:} $\cmax/\cmin$ for the 1024 ms channel vs. that for the
64 ms channel. All bursts with $\cmax/\cmin$ available in both
the 64 ms and 1024 ms channels are plotted.
Open and filled circles correspond to bursts with durations shorter
and longer than 2s respectively. Note that $\cmax/\cmin$ in the 1024 ms
channel is larger than $\cmax/\cmin$ in the 64 ms channel for almost all
the long bursts, while the short bursts tend to have larger $\cmax/\cmin$
in the 64 ms channel.

\vspace{0.3cm}
\noindent
{\bf Fig. 2:} The thick line shows the probability values $p_{K-S}$ obtained
with the Kolmogorov-Smirnov (K-S) test when the $l_{1024}$ and $s_{64}$
samples are compared versus the quantity $\log R_3$ (see eq 5). Also shown as
a thin line is $\Delta/\sigma$ vs. $\log R_3$, where $\Delta$ is defined
as the absolute difference of $\vvmax$ between the two samples, and $\sigma$
is the expected standard deviation. The dotted line corresponds to a $1 \sigma$
deviation.

\vspace{0.3cm}
\noindent
{\bf Fig. 3:} $V/V_{\rm max}$ for each burst is shown versus the burst's
intensity rank normalized by the total number. The thick and thin
lines correspond to the 113 long bursts in the 1024 ms channel (sample
$l_{1024}$) and the 40 short bursts in the 64 ms channel (sample $s_{64}$).
The K-S probability that these two samples are drawn from the same
distribution is 42\%. The probabilities that these two samples are drawn
from a uniform distribution in a Euclidean space (indicated as a dashed
diagonal line) is $< 10^{-4}$.

\end{document}